\documentclass[submission,copyright,creativecommons]{eptcs}
\providecommand{\event}{SOS 2007} 
\usepackage{breakurl}             
\usepackage{underscore}           
\usepackage{amssymb}

\usepackage{cedilleverbatim}

\usepackage{subcaption} 

\usepackage{proof}
\usepackage{alltt}

\usepackage{amsthm}
\theoremstyle{definition}
\newtheorem{definition}{Definition}[section]

\usepackage{tikz-cd}

\newcommand{\abs}[4]{{#1}\, \mathit{#2}\! : \! #3.\, #4}
\newcommand{\absu}[3]{{#1}\, \mathit{#2}.\, #3}
\mathchardef\mhyph="2D 
\newcommand{\ann}[2]{\mathit{#1}\! : \! #2}
\newcommand{\ana}[1]{\ensuremath{[\!\!(#1)\!\!]}}

\title{Efficient lambda encodings for Mendler-style coinductive types in
  Cedille}
\author{
  Christopher Jenkins \qquad Aaron Stump \qquad Larry Diehl
  \institute{The University of Iowa, Iowa City, Iowa USA}
  \email{{\string{}firstname-lastname\string}@uiowa.edu}
}
\def\titlerunning{Efficient lambda encodings for Mendler-style coinductive types
in Cedille}
\def\authorrunning{C. Jenkins, A. Stump \& L. Diehl}
\begin{document}
\maketitle

\begin{abstract}
  In the calculus of dependent lambda eliminations (CDLE), it is possible
  to define inductive datatypes via lambda encodings that feature constant-time
  destructors and a course-of-values induction scheme.
  This paper begins to address the missing derivations for the dual,
  \emph{coinductive} types.
  Our derivation utilizes new methods within CDLE, as there are seemingly fundamental
  difficulties in adapting previous known approaches for deriving inductive types.
  The lambda encodings we present implementing coinductive types
  feature constant-time constructors and a course-of-values corecursion scheme.
  Coinductive type families are also supported,
  enabling proofs for many standard coinductive properties such as
  stream bisimulation.
  All work is mechanically verified by the Cedille tool, an
  implementation of CDLE.
\end{abstract}

\section{Introduction}
Inductive (algebraic) datatypes, such as
natural numbers and lists, serve the crucial role in functional languages of
structuring the definitions of functions over (necessarily) finite objects.
To each such datatype is associated an
\emph{eliminator}, a combinator for defining provably terminating functions over
such objects, justifying structurally recursive definitions.
Less familiar is the notion
of coinductive (coalgebraic) datatypes, which serve to structure the definitions
of functions generating (possibly) infinite objects. An example of a codatatype is
the type of \emph{streams}, which are infinite lists of some element type.
Analogously, associated with each coinductive type is a \emph{generator}, a
combinator for giving definitions of such objects that are provably productive (that is, all
finite observations made of these objects are defined).

Usually, inductive and coinductive types are built-in primitives to
functional languages.
Though they are sometimes declared via the same syntactic mechanisms (as in
Haskell), in total functional languages and especially in implementations of
dependent type theories they are better kept separate.
As an example of such theory, the \emph{calculus of constructions} \cite{CH88} (CC) is
extended by the \emph{calculus of inductive constructions} \cite{We94} (CIC)
with primitive inductive datatypes each with an associated induction principle;
syntactically, they are given by a declaration listing the
constructors generating elements of the datatype and used by pattern matching and
recursion.
CIC can be further extended with primitive coinductive types, and
an elegant syntactic proposal for them comes from
Abel et al. \cite{APTS13}: giving these by a declaration listing their
\emph{destructors} and generating them via copattern matching.
The elegance of this proposal lies in its close fit with the semantic account of
coinductive types developed by Hagino \cite{Ha87}.

The \emph{calculus of dependent lambda eliminations} \cite{Stu17, Stu18b} (CDLE)
is a compact Curry-style type theory with no primitive notion of (co)inductive
datatypes.
Instead, datatypes are encoded as lambda expressions.
Historically, lambda encodings have languished due to several difficulties,
which CDLE overcomes.
Geuvers~\cite{Geu01} showed the impossibility of deriving induction
for them in second-order dependent type theory; Firsov and Stump~\cite{FS18} demonstrated how to generically
derive the induction principle for them in CDLE.
Parigot~\cite{Par89} showed that the Church-style of encoding has no better than
linear-time data accessors (such as 
predecessor for natural numbers); Firsov et al.~\cite{FBS18} showed how to use
the induction principle for a Mendler-style of encoding~\cite{Me91} to define
\emph{efficient} (constant-time) data accessors.
Furthermore, Firsov et al. \cite{FDJS18} show how to further augment this with
\emph{course-of-values} induction, an expressive scheme wherein
the inductive hypothesis can be invoked on subdata at unbounded depth.

The present paper partially addresses the gap in the foregoing account of
datatypes in CDLE by presenting a generic derivation of lambda-encoded
\emph{coinductive} data, giving analogous results for efficiency (for
\emph{constructors}) and expressivity (using \emph{course-of-values coiteration}).
What we do not show here, however, is a ``true coinduction'' principle in the
sense of ``bisimilarity implying equality'' -- indeed, a negative result
concerning \emph{Lambek's lemma} (a consequence of coinduction) indicates
coinduction in this sense is not possible for our encoding.
However, our derived codatatypes support \emph{indexed coiteration}, which we
argue suffices for implementing many standard examples of coinductive reasoning
such as showing that two streams are bisimilar.
Since efficiency for Mendler-encoded data came as a consequence of induction in
\cite{FBS18}, lacking true coinduction we cannot directly apply their method.
Instead, we use (monotone) recursive types to achieve efficiency (a technique
deployed for similar effect in e.g., \cite{Ge14, Men87,
  Lei89, Mat98b}), a feature which is itself derivable within CDLE.

\paragraph{Contributions}
Summarizing, in this paper we:
\begin{itemize}
\item generically derive Mendler-style lambda encodings for coinductive
  datatypes in CDLE using derivable monotone recursive types;
  
\item show that codata so derived forms an adequate basis for functional
  programming due to the \emph{expressive course-of-values corecursion} scheme
  it supports for generating codata and their \emph{efficient codata
    constructors} (in the sense that the run-time overhead of every use of a
  constructor is constant in the number of observations made on the codata);

\item show that our derived codata supports a proof principle in the form of
  \emph{indexed course-of-values corecursion}, which suffices for proving many
  standard examples of coinductive properties such as
  relational properties over corresponding elements of two streams;
  
\item and describe a negative result concerning \emph{Lambek's lemma} that
  indicates the impossibility of coinduction in the sense of ``bisimilarity
  implying equality'' (with respect to CDLE's built-in equality) for this
  encoding; this in turn explains the use of monotone recursive types, as the
  lack of such a coinduction principle raises fundamental difficulties in
  adapting for this encoding previous approaches for deriving efficient
  encodings for \emph{inductive} types in CDLE.
\end{itemize}

The remainder of this paper is organized as follows:
in Section \ref{sec:prelim} we review CDLE and Mendler-style corecursion
schemes;
in Section \ref{sec:generic-deriv} we detail the generic derivation of
coinductive data, discussing its strengths and the
above-mentioned negative result;
in Section \ref{sec:examples} we give concrete examples of programming with streams using
expressive corecursion schemes supported by our derived codatatypes;
in Section \ref{sec:proofs} we argue that the indexed coiteration scheme
they support enables proofs of many properties of interest for coinductive types
such as streams;
in Section \ref{sec:related} we discuss related and future work; and we conclude
in Section \ref{sec:conclusion}.
All code found in listings can be found at
\url{https://github.com/cedille/cedille-developments} and is checkable by the
Cedille tool, and implementation of CDLE available at
\url{https://github.com/cedille/cedille/releases}.

\section{Preliminaries}
\label{sec:prelim}
\subsection{CDLE}
\label{sec:cdle}
We begin with a review of the CDLE, the type theory of Cedille.
CDLE is an extension of impredicative
Curry-style (i.e., extrinsically typed) CC that
overcomes some traditional short-comings of lambda encodings in type theory
(e.g., underivability of induction for them in second-order dependent type
theory \cite{Geu01}) while maintaining a compact formal description: Cedille
Core, a minimal specification of CDLE, can be implemented in $\sim$1K Haskell
LoC \cite{Stu18c}.
CDLE accomplishes this by adding
three new type constructs: equality of untyped terms ($\{t ≃ t'\}$); the dependent
intersections ($\abs{ι}{x}{T}{T'}$) of Kopylov \cite{Kop03}; and the implicit (erased)
products ($\abs{∀}{x}{T}{T'}$) of Miquel \cite{Mi01}.
The pure term language of CDLE is
the just the untyped $\lambda$-calculus; to make type checking algorithmic, terms in
Cedille are given with typing annotations, and definitional equality of terms is modulo erasure
of annotations.
Figures \ref{fig:cdle-type-constructs} and \ref{fig:appendix-phi}
(Appendix \ref{sec:appendix-phi}) gives the typing and erasure rules for the
fragment of CDLE relevant to this paper.
In particular, the details for dependent intersections are given in the 
appendix as they are not required understanding the main results of this paper
-- they are used in Appendix \ref{sec:appendix-derived} to implement the derived
type constructors of Section \ref{sec:derived}.
A complete reference for the syntax, typing, and erasure, along with some
meta-theoretic results, can be found in \cite{Stu18b}. 

\begin{figure}
  \begin{subfigure}{1\linewidth}
    \[
      \begin{array}{c}
        \begin{array}{ccc}
          \infer{
          \Gamma \vdash \beta : \{t ≃ t\}
          }{
          \quad \mathit{FV}(t) \subseteq \mathit{dom}(\Gamma)
          }
          &
            \infer{
             \Gamma \vdash \rho\ t'\ -\ t : [t_1/x] T
            }{
            \Gamma \vdash t' : \{t_1 \simeq t_2\}
            \quad \Gamma \vdash t : [t_2/x] T
            }
          &
            \infer{
             \Gamma \vdash \delta\ T\ -\ t : T
            }{
             \Gamma \vdash t : \{ \absu{\lambda}{x}{\absu{\lambda}{y}{x}} \simeq
            \absu{\lambda}{x}{\absu{\lambda}{y}{y}} \}
            \quad \Gamma \vdash T : \star
            }
        \end{array}
        \\ \\
        \begin{array}{cc}
          \infer{
          \Gamma\vdash\abs{\Lambda}{x}{T'}{t} : \abs{\forall}{x}{T'}{T}
          }{
          \Gamma,\ann{x}{T'}\vdash t:T
          \quad x\not\in\textit{FV}(|t|)
          }
          &
            \infer{
            \Gamma\vdash t\ \mhyph t' : [t'/x]T
            }{
            \Gamma\vdash t : \abs{\forall}{x}{T'}{T}
            \quad \Gamma\vdash t' : T'
            }
        \end{array}
      \end{array}
    \]
  \end{subfigure}
  \begin{subfigure}{0.5\linewidth}
    \[
      \begin{array}{rcl}
        |\beta| & = & \absu{\lambda}{x}{x}
        \\ |\rho\ t'\ -\ t| & = & |t|
        \\ |\delta\ T\ -\ t| & = & \absu{\lambda}{x}{x}
      \end{array}
    \]
  \end{subfigure}%
  \begin{subfigure}{0.5\linewidth}
    \[
      \begin{array}{rcl}
        |\abs{\Lambda}{x}{T}{t}| & = & |t|
        \\ |t\ \mhyph t'| & = & |t|
      \end{array}
    \]
  \end{subfigure}
  \caption{A fragment of CDLE (equality and implicit products)}
  \label{fig:cdle-type-constructs}
\end{figure}

\paragraph{Equality}
$\{t_1 \simeq t_2\}$ is the type of proofs that the erasures of $t_1$ and
$t_2$ ($|t_1|$ and $|t_2|$, resp.) are equal. It is introduced with $\beta$,
which proves $\{t \simeq t\}$ for any $t$ whose free variables are declared in
the typing context.
The term $\beta$ erases to $\absu{\lambda}{x}{x}$, similar to the lambda
encoding for Leibniz equality in CC.
Combined with definitional equality,
$\beta$ proves $\{t_1 ≃ t_2\}$ for terms $t_1$ and $t_2$ whose erasures
are $\beta\eta$-convertible.

We eliminate equality proofs with $\rho\ t'\ -\ t$, a substitution principle,
and with $\delta\ T\ -\ t$, which provides the \emph{principle of explosion} for
a certain contradictory equation.
For substitution with $\rho$, if the expected type of the expression $\rho\ t'\
-\ t$ is $T[t_1/x]$, and $t'$ proves $\{t_1 ≃ t_2\}$, then $t$ is checked
against the type $T[t_2/x]$.
The expression $\rho\ t'\ -\ t$ erases to $|t|$, making equality in Cedille
\emph{proof-irrelevant} (in contrast to theories like CIC and MLTT).
For $\delta$, if $t$ proves that the lambda encoding of the Boolean for
\emph{true} is equal to the encoding for \emph{false}, $\delta\ T\ -\ t$
has type $T$, for any type $T$.
The expression $\delta\ T\ -\ t$ erases to $\absu{\lambda}{x}{x}$, so
again the proof of equality $t$ is computationally irrelevant.
As a convenience, the Cedille tool implements for $\delta$ the \emph{B\"ohm-out
  algorithm} \cite{BDPR79} so that it may be used on proofs $\{t_1 ≃
t_2\}$ for any closed and normalizing terms $t_1$ and $t_2$ whose erasures are
$\beta\eta$-inconvertible.



\paragraph{Implicit product}
$\abs{\forall}{x}{T'}{T}$ is the type of dependent functions with an
\emph{erased} (computationally irrelevant) argument of type $T'$ and a result of
type $T$.
They are introduces with an abstraction $\abs{\Lambda}{x}{T'}{t}$ where we require
that $x$ does not occur free in the erasure of $t$, permitting the erasure of
$\abs{\Lambda}{x}{T'}{t}$ to be $|t|$.
Because of this restriction on $x$, erased arguments play no computational role
in the body $t$ and thus exist solely for the purposes of typing.

Terms $t$ having an implicit product type $\abs{\forall}{x}{T'}{T}$ are
eliminated with \emph{erased application}: if $t'$ has type $T'$, then the
erased application $t\ \mhyph t'$ has type $T[t'/x]$ and erases to $|t|$.
When $x$ does not occur free in the type $T'$, we write $T ➾ T'$ for $\abs{∀}{x}{T}{T'}$,
similar to $T ➔ T'$ for non-dependent non-implicit products.

\paragraph{Inherited type constructs from CC.}
Figure \ref{fig:cdle-type-constructs} omits typing and erasure rules for the
term and type constructs of CC.
In terms, all type abstractions $\absu{\Lambda}{X}{t}$ (so, $\Lambda$ is used
for introducing polymorphic terms as well as for functions with erased term arguments)
erase to $|t|$, and all term-to-type applications $t \cdot S$ erase to $|t|$.
In types, $\forall$ also quantifies over types, $\abs{\Pi}{x}{T'}{T}$ is a (non-implicit)
dependent product, and $\lambda$ introduces a type-level function.
In code listings, we omit type arguments and annotations when Cedille can infer these.

\subsection{Derived constructs}
\label{sec:derived}
Our derivation of coinductive datatypes makes use of several constructions which
are themselves derived within CDLE.
However, due to space restrictions we do not detail their definitions,
instead choosing to present them axiomatically with type inference rules and (when
appropriate) rules for definitional equality.
These include familiar non-recursive datatypes -- \texttt{Pair} for 
pairs and \texttt{Unit} for the singleton type (Figure \ref{fig:utils}) -- all
derivable following a similar approach as described by Stump \cite{Stu18a}.
Some of the derived type constructors are more exotic (Figure
\ref{fig:derived}): \texttt{Cast} for zero-cost type coercions, \texttt{Mono}
for internalized monotonicity witnesses of type schemes, and a recursive type
former \texttt{Rec}.
We describe this latter kind, whose complete derivation in Cedille is listed in
Appendix \ref{sec:appendix-derived} (and in the code repository for this paper),
and are described at length by Jenkins and Stump in \cite{JS20}.
In order to support indexed coinductive types, the definitions in Figure
\ref{fig:derived} and through Section \ref{sec:generic-deriv} make use of
an indexing type $\ann{I}{★}$.

\begin{figure}
  \centering
  \begin{subfigure}{1.0\linewidth}
    \caption{\texttt{Cast} (zero-cost coercions)}
    \label{fig:casts}
    \[
      \begin{array}{c}
        \begin{array}{cc}
          \infer{
           \Gamma \vdash \texttt{Cast} ·I ·S ·T : ★
          }{
           \Gamma \vdash S : I ➔ ★ \quad \Gamma \vdash T : I ➔ ★
          }
          &
            \infer{
            \Gamma \vdash \texttt{elimCast}\ \mhyph c\ \mhyph i : S\ i ➔ T\ i
            }{
             \Gamma \vdash c : \texttt{Cast} ·I ·S ·T
             \quad \Gamma \vdash i : I
            }
        \end{array}
        \\ \\
        \infer{
         \Gamma \vdash \texttt{intrCast}\ \mhyph f\ \mhyph p : \texttt{Cast} ·I ·S ·T
        }{
         \Gamma \vdash f : \abs{∀}{i}{I}{A\ i ➔ B\ i}
         \quad
         \Gamma \vdash p : \abs{∀}{i}{I}{\abs{Π}{x}{S\ i}{\{f\ x ≃ x\}}}
        }
        \\ \\
        \begin{array}{cc}
          \infer{
           \Gamma \vdash \texttt{castTrans}\ \mhyph t_1\ \mhyph\ t_2 :
          \texttt{Cast} ·I ·S ·U
          }{
          \Gamma \vdash t_1 : \texttt{Cast} ·I ·S ·T
          \quad \Gamma \vdash t_2 : \texttt{Cast} ·I ·T ·U
          }
          &
            \infer{
             \Gamma \vdash \texttt{castRefl} ·I ·S : \texttt{Cast} ·I ·S ·S
            }{
             \Gamma \vdash S : I ➔ ★
            }
        \end{array}
        \\ \\
        \begin{array}{cc}
          \texttt{intrCast}\ =_{\beta}\ \absu{λ}{x}{x},
          & \texttt{elimCast}\ =_{\beta}\ \absu{λ}{x}{x}
          \\ 
          \texttt{castRefl}\ =_{\beta}\ \absu{λ}{x}{x},
          & \texttt{castTrans}\ =_{\beta}\ \absu{λ}{x}{x}
        \end{array}
        \\ \\
      \end{array}
    \]
  \end{subfigure}
  \begin{subfigure}{1.0\linewidth}
    \caption{\texttt{Mono} (internalized monotonicity predicate)}
    \label{fig:mono}
    \[
      \begin{array}{c}
        \begin{array}{cc}
          \infer{
             \Gamma \vdash \texttt{Mono} ·I ·F : ★
          }{
             \Gamma \vdash I : ★
             \quad \Gamma \vdash F : (I ➔ ★) ➔ I ➔ ★
          }
          &
            \infer{
            \Gamma \vdash \texttt{elimMono}\ \mhyph m\ \mhyph c\ \mhyph i : F ·S\
            i ➔ F ·T\ i
            }{
            \Gamma \vdash m : \texttt{Mono}·I ·F
            \quad \Gamma \vdash c : \texttt{Cast} ·I ·S ·T
            \quad \Gamma \vdash i : I
            }
        \end{array}
        \\ \\
        \infer{
           \Gamma \vdash \texttt{intrMono}\ \mhyph f : \texttt{Mono} ·I ·F
        }{
            \Gamma \vdash f : \abs{∀}{X}{I ➔ ★}{
             \abs{∀}{Y}{I ➔ ★}{
              \texttt{Cast} ·I ·X ·Y ➾ \texttt{Cast} ·I ·(F ·X) ·(F ·Y)}}
        }
        \\ \\
        \begin{array}{cc}
          \texttt{intrMono}\ =_{\beta}\ \absu{λ}{x}{x},
          & \texttt{elimMono}\ =_{\beta}\ \absu{λ}{x}{x}
        \end{array}
        \\ \\
      \end{array}
    \]
  \end{subfigure}
  \begin{subfigure}{1.0\linewidth}
    \caption{\texttt{Rec} (recursive type former)}
    \label{fig:rec}
    \[
      \begin{array}{c}
        \begin{array}{cc}
          \infer{
            \Gamma \vdash \texttt{Rec} ·I ·F : I ➔ ★
          }{
            \Gamma \vdash I : ★
            \quad \Gamma \vdash F : (I ➔ ★) ➔ I ➔ ★
          }
          &
            \infer{
              \Gamma \vdash \texttt{unroll}\ \mhyph m\ \mhyph i : \texttt{Rec} ·F\ i ➔
              F·(\texttt{Rec} ·F)\ i
            }{
              \Gamma \vdash m : \texttt{Mono} ·I
              \quad \Gamma \vdash i : I
            }
        \end{array}
        \\ \\
        \infer{
         \Gamma \vdash \texttt{roll}\ \mhyph m\ \mhyph i : F·(\texttt{Rec}·F)\ i ➔
         \texttt{Rec}·F\ i
        }{
         \Gamma \vdash m : \texttt{Mono} ·I ·F
         \quad \Gamma \vdash i : I
        }
        \\ \\
        \begin{array}{cc}
          \texttt{roll}\ =_{\beta}\ \absu{λ}{x}{x},
          & \texttt{unroll}\ =_{\beta}\ \absu{λ}{x}{x}
        \end{array}
      \end{array}
    \]
  \end{subfigure}
   \caption{Derived type constructors for monotone recursive types}
  \label{fig:derived}
\end{figure}

\paragraph{\texttt{Cast}, zero-cost type coercions.}
For type families $S$ and $T$ of kind $I ➔ ★$, \(\texttt{Cast} ·I ·S
·T\) is the type of generalized identity functions in CDLE; its formation,
introduction, and elimination rules are given in Figure \ref{fig:casts}.
This family of types
was first introduced by Firsov et al. in~\cite{FBS18}, wherein it is called
\texttt{Id}, and only the non-indexed variant is given (for the related notion
of zero-cost coercion in Haskell, see \cite{BEPW16}).
Since CDLE is Curry-style, an identity function from $S$ to $T$ might exists
even if $S$ and $T$ are inconvertible types.
Terms of type $\texttt{Cast} \cdot I \cdot S \cdot T$ are introduced with
\(\texttt{intrCast}\ \mhyph f\ \mhyph p\) (that is, with both arguments erased),
where $f$ has type $\abs{∀}{i}{I}{S\ i ➔ T\ i}$ and $p$ (of type
$\abs{\forall}{i}{I}{\abs{\Pi}{x}{S\ i}{\{f\ x \simeq 
    x\}}}$) is a proof that $f$ behaves \emph{extensionally} like the identity
function (there is no need to write the equality as \(\{f\ \mhyph i\ x \simeq
x\}\), as equality for terms is modulo erasure).
Casts are eliminated with \texttt{elimCast}: if $c$ has type $\texttt{Cast}
\cdot I \cdot S \cdot T$ and $i$ has type $I$, then  \(\texttt{elimCast}\ \mhyph
c\ \mhyph i\) is a function from $S\ i$ to $T\ i$, which (crucially) is
definitionally equal to $\absu{\lambda}{x}{x}$ (indicated by the
notation $=_{\beta}$ in the figure).
Since \texttt{elimCast} takes its cast argument erased, \texttt{Cast} is also
proof-irrelevant.
Constructs \texttt{castRefl} and \texttt{castTrans} give us that
casts form a preorder on type families (there can be at most one
identity function between any two type families) and can be 
defined in terms of \texttt{intrCast} and \texttt{elimCast}.
Both \texttt{castRefl} and \texttt{castTrans} are definitionally equal to
$\absu{λ}{x}{x}$ as well.


\paragraph{\texttt{Mono}, internalized positivity.}
Given $\ann{F}{(I ➔ ★) ➔ I ➔ ★}$, \(\texttt{Mono} ·I ·F\) (Figure
\ref{fig:mono}) is the type of proofs
that the type scheme $F$ is \emph{monotonic}.
Monotonicity for type schemes (as opposed to \emph{syntactic} positivity) as
presented here resemble the work of Matthes \cite{Mat99, Mat98b},
in which monotonicity is given by ``terms of functorial strength''.
Terms of type \(\texttt{Mono} \cdot I \cdot F\) are introduced by \texttt{intrMono}, which
takes as an argument some $f$ which (similar to the more familiar morphism mapping
of a functor in category theory), for all type families $X$ and $Y$ of kind $I ➔
★$, transforms an (erased) cast from $X$ to $Y$ to a cast from $F ·X$ to $F ·Y$.
A witness of positivity $m$ of type \(\texttt{Mono} ·I ·F\) can be eliminated
with \(\texttt{elimMono}\ \mhyph m\ \mhyph c\ \mhyph i\) to a function of type
$F \cdot S\ i \to F \cdot T\ i $, where
$c$ has type $\texttt{Cast} ·I ·S ·T$ (for some type families $S$ and $T$) and
$i$ has type $I$.
The entire expression is definitionally equal to $\absu{λ}{x}{x}$ (notice that
all arguments to \texttt{elimMono} are erased). 

\paragraph{\texttt{Rec}, a recursive type former.}
Given  $F : (I ➔ ★) ➔ I ➔ ★$,
\(\texttt{Rec} ·I ·F\) (Figure \ref{fig:rec}) is a fixedpoint of $F$ (strictly speaking, it is the
\emph{least} fixedpoint in the pre-order on type schemes induced by
\texttt{Cast}).
It is well-known that unrestricted formation and use of recursive types is
unsound when interpreting type theories under the Curry-Howard isomorphism,
and so accordingly the introduction (\texttt{roll}) and
elimination (\texttt{unroll}) of terms with recursive types requires evidence that the type
scheme $F$ whose fixedpoint was taken is monotonic.
If $m$ has type $\texttt{Mono} ·I ·F$ and $i$ has type $I$, then \(\texttt{roll}\ \mhyph
m\ \mhyph i\) is a function taking some term of type $F ·(\texttt{Rec} ·F)\ i$
and producing a term of type \(\texttt{Rec} ·F\ i\).
The situation is symmetric for \texttt{unroll}.

In type theories with primitive iso-recursive types, \texttt{roll} and
\texttt{unroll} must form an isomorphism.
Usually, definitional equality is extended with the $\beta$-law for
\texttt{Rec}, meaning \(\texttt{unroll}\ (\texttt{roll}\ t)\) reduces to $t$,
with the $\eta$-law \(\texttt{roll}\ (\texttt{unroll}\ t) = t\) possibly holding
only meta-theoretically.
The derived recursive types of CDLE differ from this in that both \texttt{roll}
and \texttt{unroll} are each themselves definitionally equivalent to
$\absu{\lambda}{x}{x}$; in this respect, recursive types in CDLE are akin to \emph{equi-recursive}
types as the terms \(\texttt{roll}\ \mhyph m\ \mhyph i\ t\) and $t$ are
definitionally equivalent (for $t$ of type \(F \cdot (\texttt{Rec} \cdot F)\
i\)), as well as the terms \(\texttt{unroll}\ \mhyph m\ \mhyph i\ t'\) and $t'$
(for $t'$ of type \(\texttt{Rec} \cdot F\ i\)).

\begin{figure}
  \centering
  \[
      \begin{array}{cc}
        \infer{
        \Gamma \vdash \texttt{Pair} ·A ·B : ★
        }{
        \Gamma \vdash A : ★
        \quad \Gamma \vdash B : ★
        }
        &
          \infer{
          \Gamma \vdash \texttt{intrPair}\ t_1\ t_2 : \texttt{Pair} ·A ·B
          }{
          \Gamma \vdash t_1 : A
          \quad \Gamma \vdash t_2 : B
          }
        \\ \\
        \infer{
         \Gamma \vdash \texttt{fst}\ p : A
        }{
         \Gamma \vdash p : \texttt{Pair} ·A ·B
        }
        &
          \infer{
          \Gamma \vdash \texttt{snd}\ p : B
          }{
          \Gamma \vdash p : \texttt{Pair} ·A ·B
          }
        \\ \\
        \infer{\Gamma \vdash \texttt{Unit} : ★}{}
        &
          \infer{\Gamma \vdash \texttt{unit} : \texttt{Unit}}{}
    \end{array}
  \]
  \caption{Derived datatypes: pairs and the unitary type}
  \label{fig:utils}
\end{figure}

\subsection{Coalgebras and coiteration schemes}
\label{sec:coalgebras}

In category theory, coinductive datatypes are understood as \emph{final
  $F$-coalgebras} \cite{Ha87}.
Our generic derivation rests upon this
understanding, but in order to support efficient constructors and
course-of-values corecursion we find it convenient to have our semantic
account begin with \emph{Mendler-style} $F$-coalgebras.
The dual notion, the Mendler-style $F$-algebra, is discussed in more depth by
Uustalu and Vene~\cite{UV99} and by Vene \cite{Ven00}; the following categorical
account is a straightforward adaptation of this description for $F$-coalgebras.

\begin{definition}[Ordinary $F$-coalgebras]
  Assuming $F : \mathcal{C} \to \mathcal{C}$ is a functor, the usual definition of
  an $F$-coalgebra is a pair $(X,\phi)$ where $X$ is an object (e.g., a type)
  called the \emph{carrier} of the algebra and $\phi: X \to F\ X$ is a morphism (e.g., a
  function) from $X$ to $F\ X$ called the \emph{action} of the coalgebra.
\end{definition}

Translated to type theory, the actions of $F$-coalgebras are expressed by the
family of types
\[
  \abs{\lambda}{X}{\star}{X \to F \cdot X}
\]

\begin{definition}(Mendler-style $F$-coalgebras)
  A \emph{Mendler-style
    $F$-coalgebra} is a pair $(X,\Phi)$ where the carrier $X$ is an object and the
  action $\Phi$ is a natural transformation (e.g., a polymorphic function) that, for
  every object $R$ in $\mathcal{C}$, maps elements of $\mathcal{C}(X,R)$ (the set of
  morphisms from $X$ to $R$) to elements of $\mathcal{C}(X,F\ R)$.
\end{definition}

Translated to type theory, the actions of Mendler $F$-coalgebras are
expressed by the family of types:
\[
  \texttt{CoAlgM} = \abs{λ}{X}{★}{\abs{∀}{R}{★}{(X \to R) \to X \to F ·R}}
\]
\noindent In functional programming, the operational reading of the type $X \to
F\cdot X$ of
an action of an ordinary $F$-coalgebra $(X,\phi)$ is that it is a function
taking some ``seed value'' of type $X$ and producing some ``structure'' of type
$F\cdot X$.
For an action of a Mendler $F$-coalgebra polymorphic in $R$, it instead
produces a structure of type $F\cdot R$ given both a seed value of type $X$ and a function
for transforming such seed values to values of type $R$.

\begin{definition}[Final Mendler $F$-coalgebras]
  A \emph{final} Mendler-style $F$-coalgebra $(\nu F, \mathit{out}^F)$ is one such
  that for every other Mendler-style $F$-coalgebra $(X,\Phi)$, there exists a
  unique morphism $\ana{\Phi} : X \to \nu F$ (called the \emph{Mendler anamorphism} of
  $\Phi$) such that
  
  \[\mathit{out}^F_{\nu F}(\mathit{id}) \circ \ana{\Phi} = \Phi_{\nu
      F}(\ana{\Phi}) = (F\ \ana{\Phi}) \circ \Phi_X(\mathit{id})\]
  
  \noindent where subscripts on natural transformations indicating
  component selection (e.g., polymorphic instantiation) and $F\ \ana{\Phi} : F\ X
  \to F \nu F$
  denotes the functorial lifting of $\ana{\Phi}$ to a morphism from $F\ X$ to $F\
  \nu F$. This condition can be alternatively be stated as saying that, for any
  Mendler $F$-coalgebras $(X,\Phi)$, $\ana{\Phi}$ is the unique morphism making
  the diagram below commute:
  \[
    \begin{tikzcd}[column sep=huge, row sep=large]
      X
      \arrow[dr,"\Phi_{\nu F}(\ana{\Phi})" sloped]
      \arrow[r,"\Phi_X(\mathit{id})"]
      \arrow[d,"\ana{\Phi}"]
      & F\ X
      \arrow[d,"F\ \ana{\Phi}"]
      \\
      \nu F
      \arrow[r,"\mathit{out}^F_{\nu F}(\mathit{id})" below]
      &
      F\ \nu F
    \end{tikzcd}
  \]
\end{definition}

In type theory, the carrier $\nu F$ of the final Mendler $F$-coalgebra can be
given as:

\[
  \texttt{Nu} = \abs{∃}{X}{★}{X \times \texttt{CoAlgM} ·X}
\]
\noindent or encoded as
$\abs{∀}{Y}{★}{(\abs{∀}{X}{★}{X ➔ \texttt{CoAlgM} ·X ➔ Y}) ➔ Y}$ if existentials
and products are unavailable.

Intuitively, $\mathit{out}_{\nu F}^F$ is the generic destructor for the coinductive
datatype $\nu F$ with pattern functor $F$, and for any Mendler $F$-coalgebra $(X,\Phi)$,
$\ana{\Phi}$ is a generator for $\nu F$.
The equations express how to \emph{compute} the observations
$\mathit{out}^F_{\nu F}(\mathit{id})$ for codata generated with $\ana{\Phi}$:
these are given by simply calling $\Phi$ with the generator $\ana{\Phi}$, or
equivalently by first calling $\Phi$ with a trivial generator $\mathit{id}_X$,
then mapping over the resulting $F\ X$ structure with the generator $\ana{\Phi}$.

\paragraph{Mendler-style coiteration}
The Mendler-style anamorphism translates to a \emph{coiteration} scheme $\mathit{coit}$
for codata, whose typing and computation rules are given below (with
$\mathit{out} : \nu F \to F\ \nu F$ below corresponds to
$\mathit{out}^F_{\nu F}(\mathit{id})$ above).

\[
  \begin{array}{lr}
    \infer{
     \mathit{coit}\ a : X \to \nu F
    }{
    a : \abs{\forall}{R}{\star}{(X \to R) \to X \to F \cdot R}
    \quad x : X
    }
    &
      \mathit{out}\ (\mathit{coit}\ a\ x) \twoheadrightarrow a\ (\mathit{coit}\ a)\ x
  \end{array}
\]
\noindent Read operationally, the type of $a$ suggests that it will,
from a seed value of type $X$ and function for making coiterative calls $X \to
R$ (where $R$ is universally quantified over), construct an $F$-collection of
$R$ subdata, corresponding to one additional observation 
that can later be made on the codata being generated.

\paragraph{Mendler-style corecursion}
The Mendler-style \emph{corecursion} scheme (categorically, the \emph{apomorphism}) can be
described by equipping the coalgebra action $a$ 
with an additional argument of type $\nu F \to R$, supporting an alternative method for constructing
codata -- a way of ``short-cutting'' coiteration by injecting codata directly
into the abstracted type $R$ rather than generating it from values of type $X$.
We expect that the type argument of $a$ will always be instantiated to $\nu F$,
and thus that $\absu{\lambda}{x}{x}$ can be given for this additional argument:

\[
  \begin{array}{lr}
    \infer{
     \mathit{corec}\ a : X \to \nu F
    }{
     a : \abs{\forall}{R}{\star}{(\nu F \to R) \to (X \to R) \to X \to F \cdot R}
    }
    &
      \mathit{out}\ (\mathit{corec}\ a\ x)\ \twoheadrightarrow a\ (\absu{\lambda}{x}{x})\
      (\mathit{corec}\ a)\ x
  \end{array}
\]

\paragraph{Mendler-style course-of-values coiteration}
The Mendler-style \emph{course-of-values coiteration} scheme (categorically, the
\emph{futumorphism}) can be given by equipping $a$ with an
argument of type $F ·R \to R$, another alternative to coiteration for
constructing codata that takes an $F$-collection of $R$ subdata.
This additional argument enables $a$ to construct an arbitrary number of
observations for the codata being generated \emph{in addition to} those produced
by coiteration; put another way, the $F ·R$ result that $a$ produces may have
some $R$ sub-component built using this additional argument from another $F ·R$
expression, which itself would be later accessed by making (at least) two
observations on the generated codata.
We expect that $\mathit{out}^{-1} : F ·\nu F ➔ \nu F$ (that is, the generic codata
constructor) can be given for this argument:

\[
  \begin{array}{lr}
    \infer{
    \mathit{cov}\ a : X \to \nu F
    }{
    a : \abs{\forall}{R}{\star}{(F \cdot R \to R) \to (X \to R) \to X \to F \cdot R}
    }
    &
      \mathit{out}\ (\mathit{cov}\ a\ x)\ \twoheadrightarrow
       a\ \mathit{out}^{-1}\ (\mathit{cov}\ a)\ x
  \end{array}
\]

A categorical account of the corecursion scheme for ordinary $F$-coalgebras is
given by Geuvers \cite{Ge92}, and an account of the
recursion scheme and course-of-values iteration scheme (duals to the corecursion
scheme and course-of-values coiteration scheme) for Mendler-style $F$-algebras is given by Vene
\cite{Ven00}.
For understanding the derivation of coinductive data in the next section, which
supports a combination of the corecursion and course-of-values coiteration
schemes which we call \emph{course-of-values corecursion}, the above
type-theoretic account suffices.
\section{Generic lambda encoding for codata}
\label{sec:generic-deriv}

Figures \ref{fig:nu-1}, \ref{fig:nu-2}, and \ref{fig:nu-3} give the complete
derivation of coinductive data in Cedille.
This derivation is \emph{generic}, in that it works for any monotone type scheme
$F$ of kind $(I \to \star) \to I \to \star$ (where monotonicity is expressed as
evidence $cm$ of type $\texttt{Mono} ·I ·F$).
$I$, $F$, and $\mathit{cm}$ are all module
parameters to the derivation, and the curly braces around $\mathit{cm}$ indicate
that it is an \emph{erased} module parameter (as suggested by the type inference and
definitional equal rules in Figure \ref{fig:mono}, the type of monotonicity
witnesses is \emph{proof-irrelevant}).
We walk through code listings of these figures in detail.

\begin{figure}
  \centering
  \small
\begin{verbatim}
import utils.

module nu (I: ★) (F: (I ➔ ★) ➔ I ➔ ★) {cm: Mono ·I ·F}.

CoAlgM : (I ➔ ★) ➔ (I ➔ ★) ➔ ★
= λ X: I ➔ ★. λ C: I ➔ ★.
  ∀ R: I ➔ ★. Cast ·I ·C ·R ➾ (∀ i: I. F ·R i ➔ R i) ➔
  Π ch: (∀ i: I. X i ➔ R i). ∀ i: I. X i ➔ F ·R i.

NuF : (I ➔ ★) ➔ I ➔ ★
= λ C: I ➔ ★. λ i: I. ∀ Y: I ➔ ★. (∀ X: I ➔ ★. X i ➔ CoAlgM ·C ·X ➔ Y i) ➔ Y i.

Nu : I ➔ ★ = Rec ·I ·NuF.
\end{verbatim}
  \caption{Generic lambda encoding for codata (part 1)}
  \label{fig:nu-1}
\end{figure}

\paragraph{\texttt{CoAlgM}.}
In Figure \ref{fig:nu-1}, our variant Mendler-style $F$-coalgebra,
\texttt{CoAlgM}, generalizes the description given in Section
\ref{sec:coalgebras} to type families \texttt{X: I ➔ ★}.
\texttt{CoAlgM} takes an additional type family argument \texttt{C}, and
describes a family of polymorphic functions whose arguments facilitate a
combined course-of-values corecursion scheme for codata.
The variable \texttt{C} stands in for occurrences of the generic codatatype
\texttt{Nu} -- which itself recursively defined in terms of \texttt{CoAlgM}.
In the body of the definition, the first additional argument is a \emph{cast}
from \texttt{C} to the quantified \texttt{R}, enabling \emph{corecursion}
by allowing us to produce (for any \verb|i| of type \verb|I|) some
term of type \texttt{R i} from some pre-existing codata of type \texttt{C i},
rather than via \texttt{ch}.
The second argument is an \emph{abstract constructor}: given some collection of
subdata of type \texttt{F ·R i} (for any \texttt{i}), it builds a value of type
\texttt{R i}.
This additional argument enables \emph{course-of-values 
coiteration}: for each single step of codata generation, an \emph{arbitrary} number of
observations that will be made of the codata can be constructed.
The remaining arguments are the same as for ordinary Mendler-style $F$-coalgebras:
\texttt{ch} is the handle for making coiterative calls, and the
argument of type \texttt{X i} is the value from which we are to coiteratively construct the
result of type \texttt{F ·R i}.

\paragraph{\texttt{NuF} and \texttt{Nu}.}
Type family \texttt{NuF} is defined using the standard type for lambda encodings
of existentials and products, and its body is more naturally read as \texttt{∃
  X: I ➔ ★. X i $\times$ CoAlgM ·C ·X}.
This is similar to the standard definition of the greatest fixpoint of
\texttt{F} (Section \ref{sec:coalgebras}, see also \cite{Wad90}).
\texttt{NuF} is parameterized
by a type family \texttt{C: I ➔ ★} standing in for recursive reference to a
fixpoint of itself. That fixpoint is \texttt{Nu}, defined as \texttt{Rec ·I ·NuF}.

\begin{figure}
  \centering
  \small
\begin{verbatim}
monoCoAlgM : ∀ X: I ➔ ★. Mono ·I ·(λ C: I ➔ ★. λ i: I. CoAlgM ·C ·X) = Λ X.
  intrMono -(Λ C1. Λ C2. Λ c.
    intrCast
      -(Λ i. λ coa. Λ R. Λ c'. coa -(castTrans -c -c'))
      -(Λ i. λ coa. β)) .

monoNuF : Mono ·I ·NuF = <..>

nuRoll   : ∀ i: I. NuF ·Nu i ➔ Nu i = Λ i. roll   -monoNuF -i .
nuUnroll : ∀ i: I. Nu i ➔ NuF ·Nu i = Λ i. unroll -monoNuF -i .
\end{verbatim}
  \caption{Generic lambda encoding for codata (part 2)}
  \label{fig:nu-2}
\end{figure}

\paragraph{\texttt{monoCoAlgM}, \texttt{monoNuF}, \texttt{nuRoll}, and \texttt{nuUnroll}}
As discussed earlier, recursive types must be restricted in some fashion to
positive (monotonic) type schemes only.
The term \texttt{monoNuF} in Figure \ref{fig:nu-2} (definition omitted, indicated by
\texttt{<..>}) proves that the type scheme \texttt{NuF} 
is positive; it uses evidence \texttt{monoCoAlgM} that \texttt{CoAlgM} is
positive in its first argument.
To show that a type scheme is positive (\texttt{Mono}, Section
\ref{sec:derived}), we use \texttt{intrMono} and may then assume two type
families \texttt{C1} and \texttt{C2} and a cast \texttt{c} between them.
Aside from \texttt{monoCoAlgM}, in code listings we omit monotonicity proofs as
once the general principle is understood these can be quite tedious to work through.

For \texttt{monoCoAlgM}, the goal is to show there is a cast from
\texttt{CoAlgM ·C1 ·X} to \texttt{CoAlgM ·C2 ·X} (for any \texttt{X} of kind \texttt{I ➔ ★}).
This is done with \texttt{intrCast} (also in Section \ref{sec:derived}), whose
first argument is simply a function of this type and whose second is a proof that
this function is extensionally equal to $\absu{λ}{x}{x}$.
For the functional argument, it is simply a matter of deriving a \texttt{Cast ·I
·C1 ·R} (needed for the assumed \texttt{coa} of type \texttt{CoAlgM ·C1 ·R}) from the given
\texttt{c'} of type \texttt{Cast ·C2 ·R} and \texttt{c} of type \texttt{Cast ·C1 ·C2}, which we have by
\texttt{castTrans}, the composition operator for casts.
This new cast is given to \texttt{coa} as an erased argument, so the function
simply erases to \verb|λ coa. coa|, and therefore the proof that this is
extensionally equal to $\absu{λ}{x}{x}$ is trivial (\verb|λ coa. β|),
since it is \emph{intensionally} equal to it.

  With these proofs that \texttt{NuF} is a positive type scheme, we can define
the rolling and unrolling operations \texttt{nuRoll} and \texttt{nuUnroll} for
the recursive type family \texttt{Nu}.

\begin{figure}
  \centering
  \small
\begin{verbatim}
unfoldM : ∀ X: I ➔ ★. CoAlgM ·Nu ·X ➔ ∀ i: I. X i ➔ Nu i
= Λ X. λ coa. Λ i. λ x. nuRoll -i (Λ Y. λ f. f x coa) .

inM : ∀ i: I. F ·Nu i ➔ Nu i
= unfoldM ·(F ·Nu) (Λ R. Λ c. λ v. λ ch. Λ i. λ x. elimMono -cm -c -i x) .

outM : ∀ i: I. Nu i ➔ F ·Nu i
= Λ i. λ co.
  nuUnroll -i co ·(F ·Nu)
    (Λ X. λ x. λ coa. coa -(castRefl ·I ·Nu) inM (unfoldM coa) -i x) .
\end{verbatim}
  \caption{Generic lambda encoding for codata (part 3)}
  \label{fig:nu-3}
\end{figure}

\paragraph{\texttt{unfoldM} and \texttt{inM}.}
We now discuss the definitions given in Figure \ref{fig:nu-3}.
Function \texttt{unfoldM} is the generator for our Mendler-style codata.
For any type family \texttt{X} of kind \texttt{I ➔ ★}, given some \verb|coa| of
type \verb|CoAlgM ·Nu ·X| and some \verb|x| of type \verb|X i| (for arbitrary
\verb|i| of type \verb|I|), we generate a term of
type \texttt{Nu i} by first invoking \texttt{nuRoll}, which obligates us to
produce some argument of type \texttt{NuF ·Nu i}.
Unfolding the definition of \texttt{NuF}, we provide for this an encoded
existential: a function
polymorphic over a type family \texttt{Y} of kind \texttt{I ➔ ★} taking
another function \texttt{f} of type \verb|∀ X: I ➔ ★. X i ➔ CoAlgM ·Nu ·X ➔ Y i|
and applying \texttt{f} to the given \texttt{x} and \texttt{coa}.

Our generic constructor \texttt{inM} is defined in terms of \texttt{unfoldM}.
The coalgebra given to \texttt{unfoldM} ignores its argument \texttt{ch} and instead casts (using
\texttt{cm}, the module parameter that proves monotonicity of \texttt{F}, and
\texttt{c}, the coalgebra's assumed \texttt{Cast ·I ·Nu ·R}) the
assumed \verb|x| of type \verb|F ·(Nu ·F)| to the type \texttt{F ·R}.
Avoidance of \texttt{ch}, the coalgebra's handle for coiterative calls, ensures
that \texttt{inM} is \emph{efficient} -- future observations of the codata
constructed with \texttt{inM} will not needlessly rebuild the sub-components of
the codata with which it was constructed.
This discussion is made more concrete in Section \ref{sec:lambek}.

\paragraph{\texttt{outM}.}
Finally, \texttt{outM} is our generic destructor for codata.
Given some \verb|co| of type \verb|Nu i|, we use the unrolling
operation for recursive type \texttt{Nu} on \texttt{co} to produce a term of
type \texttt{NuF ·Nu i}. Unfolding the definition of \texttt{NuF}, we provide
the resulting expression a function which assumes a type family \texttt{X} of
kind \texttt{I ➔ ★}, a term \verb|x| of type \verb|X i|, and a coalgebra
\verb|coa| of type  \texttt{CoAlgM ·(Nu ·F) ·X}.
In the body of this given function, we eliminate \texttt{coa} by instantiating its
type argument to \texttt{Nu ·F} and giving it a cast
from \texttt{Nu ·F} to \texttt{Nu ·F} (using \texttt{castRefl}), the constructor
\texttt{inM}, a handle for making coiterative calls \texttt{unfoldM coa}, and
the assumed \texttt{x}.
We can show in Cedille that the computation rules expected of our derived
codatatype \texttt{Nu} and generator \texttt{unfoldM} (Section
\ref{sec:coalgebras}) hold by $\beta$-equivalence (the cast argument, corresponding
to $\absu{λ}{x}{x}$ in the computation rule for the corecursion scheme, is
erased; \texttt{inM} corresponds to $\mathit{out}^{-1}$):

{
\begin{verbatim}
reduce : ∀ X: I ➔ ★. Π coa: CoAlgM ·Nu ·X. ∀ i: I. Π x: X i.
  { outM (unfoldM coa x) ≃ coa inM (unfoldM coa) x }
= Λ X. λ coa. Λ i. λ x. β .
\end{verbatim}
}
\subsection{\emph{Lambek's lemma}}
\label{sec:lambek}

In following sections, we shall demonstrate that our generic coinductive
datatype is adequate for both expressive functional programming 
and for giving proofs of many standard coinductive properties such as bisimulation.
Our justification for the first claim rests both on the \emph{expressivity} allowed to
programmers in defining functions producing codata (using a course-of-values
corecursion scheme) and the \emph{efficiency} of the functions so defined.
This efficiency is due to our constructor \texttt{inM} being an operation
incurring run-time overhead that is constant in the number of observations
(destructions) made on the codata.
In Cedille, we can show directly that destructing an arbitrary
constructed value produces the original collection of subdata from which the
value was constructed in a constant number of $\beta$-reductions.

{
\begin{verbatim}
lambek1 : ∀ i: I. Π xs: F ·Nu i. {outM (inM xs) ≃ xs} = Λ i. λ xs. β.
\end{verbatim}
  \normalsize
}
The name of this proof comes from \emph{Lambek's lemma} \cite{Lam68}, which
proves that (the actions of) final coalgebras are isomorphisms.
Specifically, for any functor $F$, Lambek's lemma for final (ordinary)
$F$-coalgebra $(\nu F, \mathit{out}^F)$ states that there exists an inverse
$\mathit{in}^F : F\ \nu F \to \nu F$ such that $\mathit{out}^F \circ
\mathit{in}^F = \mathit{id}_{F\ \nu F}$ and $\mathit{in}^F \circ \mathit{out}^F
= \mathit{id}_{\nu F}$.
Lambek's lemma is a consequence of the uniqueness of the anamorphism, which is
itself a category-theoretic expression of coinduction:
the usual definition of $\mathit{in}^F$ is given by the
generation of codata using the anamorphism of $F\ \mathit{out}^F : F\
\nu F \to F (F\ \nu F)$ to coiteratively rebuild its observations.
Our derivation of Mendler-style coinductive data restricts $F$ in such a way
that functorial lifting is only defined for casts ($F$ is assumed to be
\texttt{Mono}, not a functor), and our alternative definition for
$\mathit{in}^F$ (\texttt{inM}) avoids this needless rebuilding, and so one
direction of Lambek's lemma holds trivially.

Proving the Lambek's lemma in the form $\mathit{out}^F \circ
\mathit{in}^F = \mathit{id}_{F\ \nu F}$ would require \emph{functional extensionality},
but this in itself does not necessitate that the equality implied to hold
for corresponding elements in images of $\mathit{out}^F \circ \mathit{in}^F$ and
$\mathit{id}_{F\ \nu F}$ must be extensional.
Indeed, for derivations of \emph{inductive} datatypes in Cedille both directions
of Lambek's lemma need only \emph{intensional} equality.
We now show that proof of the other direction of Lambek's lemma for our derived
coinductive data appears to truly require an \emph{extensional} equality
type by giving a counter-example in Figure \ref{fig:lambek} with respect to
Cedille's currently intensional equality type. 

\begin{figure}
  \centering
  \small
\begin{verbatim}
import utils.
import nu.

module lambek.

TF : (Unit ➔ ★) ➔ (Unit ➔ ★) = λ X: Unit ➔ ★. X.
monoTF : Mono ·Unit ·TF = <..>

T : Unit ➔ ★ = Nu ·Unit ·TF monoTF.
TCoAlgM : ★ ➔ ★ = λ X: ★. CoAlgM ·Unit ·TF monoTF ·T ·(λ _: Unit. X).

tcoa : TCoAlgM ·Unit = Λ R. Λ c. λ v. λ ch. Λ i. λ x. ch -i x.
t : T unit = unfoldM -monoTF ·(λ _: Unit. Unit) tcoa -unit unit .

noLambek2 : {inM (outM t) ≃ t} ➔ ∀ X: ★. X = λ eq. Λ X. δ X - eq.
\end{verbatim}
  \caption{Violation of \emph{Lambek's lemma}}
  \label{fig:lambek}
\end{figure}

\paragraph{\texttt{TF}, \texttt{tcoa}, and \texttt{t}.}
Type scheme \texttt{TF} is the signature for \texttt{T}, and simply maps any
type family \texttt{X} of kind \texttt{Unit ➔ ★} to itself (recall that \texttt{Unit} is the
unitary type, Figure \ref{fig:utils}); our generic development requires that
every signature be indexed, so we provide \texttt{Unit} as a ``dummy'').
\texttt{TF}  is obviously monotonic, so the definition of the proof of this fact
\texttt{monoTF} is omitted (indicated by \texttt{<..>}).
This given, we define coinductive datatype \texttt{T} as the greatest fixpoint
of \texttt{TF}, and for convenience define the type family \texttt{TCoAlgM} of
\texttt{TF}-coalgebras.
Term \texttt{t} of type \texttt{T unit} is generated via \texttt{unfoldM} using
\texttt{unit} as the ``seed'' and a \texttt{TF}-coalgebra \texttt{tcoa} which
simply makes a corecursive
call with  \texttt{ch} of type \texttt{∀ i:} \texttt{Unit.} \texttt{Unit ➔ R i} to produce a
result of type \texttt{TF ·R i} (convertible with the return type \texttt{R i}
of the expression \texttt{ch -i x}).

\paragraph{\texttt{noLambek2}.}
To see why the final proof \texttt{noLambek2} holds, it is useful to rewrite the
two sides of the assumed equation \texttt{eq} to $\beta$-equivalent expressions:

{\small%
\begin{verbatim}
{ λ g. g (λ f. f unit tcoa) (λ v. λ ch. λ x. x) ≃ λ f. f unit tcoa }
\end{verbatim}
}
The right-hand side of the equation (corresponding to \texttt{t}) is a lambda
encoding of a pair consisting of a seed value \texttt{unit} and generator
\texttt{tcoa}.
However, the pair on the left-hand side has a seed value which itself is the
very same pair on the right-hand side, and a generator which simply returns the
seed value as-is.
Though we understand that these two expressions are \emph{extensionally} equal (that
is, treated as black-boxed terms of type \verb|T unit| they produce the same observations), in Cedille
the current built-in equality type is \emph{intensional}.
Indeed, $\delta$ (Section \ref{sec:cdle}) makes it \emph{anti-extensional},
though we know of no fundamental reason why CDLE could not be extended with a more
general extensional equality type.
Additionally (and as suggested by a reviewer for the draft version of this
paper), we conjecture that a reformulation of Lambek's lemma in terms of the
equivalence relation for existential types given by Reynold's relational
parametricity (such as undertaken by Pitts in \cite{Pit98}) should be provable.
In CDLE, dependent intersections can be used to equip lambda encodings with
a proof principle for parametricity in precisely the same way used to equip
encodings of inductive types with an induction principle \cite{Stu18a}; we
leave this as future work.

As we will see in Section \ref{sec:proofs}, the foregoing negative result
does not impact the ability to give proofs for many standard coinductive
properties for this encoding of codata, such as showing stream bisimilarity or
other relational properties; for many use cases, indexed coiteration suffices.
What this result indicates is currently impossible for our encoding is proving
``true coinduction'' in the sense of bisimilarity implying equality (since if this
we had this, we would be able to prove Lambek's lemma).

\section{Functional programming with streams}
\label{sec:examples}
In this section, we give a few examples of programming with lambda-encoded
codata (specifically, streams), emphasizing the expressivity the
course-of-values corecursion scheme available to programmers.
After defining the stream codatatype by
instantiating the parameters of our generic development (Figure \ref{fig:streamf}), we show how to
define: the mapping of a function over the elements of a stream, an example of
\emph{coiteration}; the mapping of a function 
over just the head element of a 
stream, an example of \emph{corecursion}; and the
pairwise exchange of elements of a stream, an example
of \emph{course-of-values coiteration}.
These example functions, given in Figure \ref{fig:progs}, also appear in Vene's
thesis \cite{Ven00}, though there they 
are given using the traditional account of coinductive types as final
$F$-algebras (not final Mendler-style ones).

\begin{figure}[t]
  \centering
  \small
\begin{verbatim}
import utils.
module examples/streamf (A: ★).

StreamF : (Unit ➔ ★) ➔ Unit ➔ ★ = λ R: Unit ➔ ★. λ u: Unit. Pair ·A ·(R u).

monoStreamF : Mono ·Unit ·StreamF = <..>

import nu/nu ·Unit ·StreamF -monoStreamF.

Stream : ★ = Nu unit.
StreamCoAlg : ★ ➔ ★ = λ X: ★. CoAlgM ·(λ _: Unit. Stream) ·(λ _: Unit. X).

head : Stream ➔ A = λ xs. fst (outM -unit xs) .
tail : Stream ➔ Stream = λ xs. snd (outM -unit xs) .

unfoldStream : ∀ X: ★. StreamCoAlg ·X ➔ X ➔ Stream = <..>
\end{verbatim}
  \caption{Definition of streams}
  \label{fig:streamf}
\end{figure}

\paragraph{Definition of streams.}
In Figure \ref{fig:streamf}, \texttt{StreamF} is the signature for streams of
elements of type \texttt{A} (where \texttt{A} is a module parameter; it is
implicitly quantified over in all definitions in the figure), defined in
terms of \texttt{Pair} (Figure \ref{fig:utils}); the definition of \texttt{StreamF} is more
recognizable in the form $R \mapsto A \times R$. Term \texttt{monoStreamF} is a
witness to the fact that \texttt{StreamF} is positive (definition omitted,
indicated by \texttt{<..>}).

The import of module \texttt{nu/nu} instantiates that module's parameters with
index type \texttt{Unit}, signature functor \texttt{StreamF}, and positivity proof
\texttt{monoStreamF}.
Type \texttt{Stream} is the greatest fixpoint of \texttt{StreamF} with a fixed
index of \texttt{unit}.
Destructors \texttt{head} and \texttt{tail} are defined in terms of the generic
destructor \texttt{out} and pair projections \texttt{fst} and \texttt{snd}.
The generator \texttt{unfoldStream} has a first argument of type
\texttt{StreamCoAlg ·X}, which specializes 
\texttt{CoAlgM} such that the type families for the codatatype and carrier
ignore their \texttt{Unit} indices.
The body of the definition of \texttt{unfoldStream} is omitted, as it is
somewhat cluttered by type annotations to replace the assumed \texttt{i:} \texttt{Unit} with the constant
\texttt{unit} in the type of the coalgebra given to \texttt{unfoldM}.
We define this function so that the following examples may use it, and not be themselves so cluttered.

\begin{figure}
  \centering
  \small
\begin{verbatim}
map : ∀ A: ★. ∀ B: ★. (A ➔ B) ➔ Stream ·A ➔ Stream ·B
= Λ A. Λ B. λ f.
  unfoldStream ·B ·(Stream ·A)
    (Λ R. Λ c. λ v. λ map. Λ i. λ xs.
     intrPair (f (head xs)) (map -i (tail xs))) .

mapHd : ∀ A: ★. (A ➔ A) ➔ Stream ·A ➔ Stream ·A = Λ A. λ f.
  unfoldStream ·A ·(Stream ·A)
    (Λ R. Λ c. λ v. λ mapHd. Λ i. λ xs.
     intrPair (f (head xs)) (elimCast -c -i (tail xs))) .

exch : ∀ A: ★. Stream ·A ➔ Stream ·A = Λ A.
  unfoldStream ·A ·(Stream ·A)
    (Λ R. Λ c. λ v. λ exch. Λ i. λ xs.
       [hd1 : A = head (tail xs)]
     - [hd2 : A = head xs]
     - [tl2 : R i = exch -i (tail (tail xs))]
     - intrPair hd1 (v -i (intrPair hd2 tl))) .
\end{verbatim}
  \caption{Programming with the coiteration, corecursion, and course-of-values coiteration schemes}
  \label{fig:progs}
\end{figure}

\paragraph{\texttt{map}.}
Our first function \texttt{map} is an example of \emph{coiteration} over
streams.
First, note that this and following functions occur in a different module than
the definition of \texttt{Stream}, so now the element type must be given
explicitly both for \texttt{Stream} and \texttt{unfoldStream}.
With a function \texttt{f} of type \texttt{A ➔ B}, in the body of \texttt{map}
we generate using \texttt{unfoldStream} a stream with elements of type
\texttt{B} from: a coalgebra wherein \texttt{map} (of type
\verb|∀ i: Unit. Stream ·A ➔ R i|) can be used for corecursive calls; and a seed
value \verb|xs| of type \verb|Stream ·A|.
In the body of the given coalgebra, we construct a \texttt{StreamF ·A} by
supplying for the head \texttt{f (head xs)} and for the tail \texttt{map -i
  (tail xs)} (of type \verb|R i|).

In a high-level surface language supporting \emph{copatterns} \cite{APTS13}, the
definition of \texttt{map} might look like:
{%
\small%
\begin{verbatim}
head (map f xs) = f (head xs)
tail (map f xs) = map f (tail xs)
\end{verbatim}%
}

\paragraph{\texttt{mapHd}.}
Function \texttt{mapHd} is an example of \emph{corecursion}.
Recall that the corecursion scheme supports an alternative for generating
codata: it may be given \emph{directly} in a coalgebra, rather than being
generated from coiteration.
This mirrors the situation for \emph{iteration} (where predecessors must be
iteratively rebuilt) and \emph{recursion} (where they are available 
``for free'') for inductive types.
In the definition, with function \texttt{f} of type \texttt{A ➔ A}, we produce
the desired stream by providing for its head \texttt{f (head xs)} and for its
tail \texttt{tail xs}, after coercing its type to the desired \texttt{R i} using
the given cast \texttt{c} from \texttt{Stream ·A} to \verb|R i|.

Using a high-level surface language, the
definition of \texttt{mapHd} might look like:
{%
\small%
\begin{verbatim}
head (mapHd f xs) = f (head xs)
tail (mapHd f xs) = tail xs
\end{verbatim}%
}

\noindent For comparison, the standard definition of corecursion from
coiteration requires ``cotupling'' (using a sum type as the generating value of
the codata).
In the same high-level pseudo-code, \texttt{mapHd} defined in this way would be more
tedious and error-prone, and result in run-time overhead linear in the number of
observations made on the codata:

{
  \small
\begin{verbatim}
mapHd' f xs = go (in1 xs) where
  go : Sum ·(Stream ·A) ·(Stream ·A) ➔ Stream ·A
  head (go (in1 ys)) = f (head ys)
  head (go (in2 ys)) = head ys
  tail (go (in1 ys)) = go (in2 (tail ys))
  tail (go (in2 ys)) = go (in2 (tail ys))
\end{verbatim}
}%

\paragraph{\texttt{exch}.}
Our final example \texttt{exch}, a function which swaps the positions of
every two elements of a stream, demonstrates the \emph{course-of-values coiteration} scheme,
wherein each step of
generating codata may describe an arbitrary number of the observations
made of that codata
(pairwise exchange only requires the construction of one additional
observation at each step).
In the coalgebra given in the definition of \texttt{exch}, we make three local
definitions (using Cedille's syntax \([x : T = t_1] - t_2\), to be read
\(\texttt{let}\ x : T = t_1\ \texttt{in}\ t_2\); we use an additional space to
distinguish hyphens used for local definitions
from hyphens used in erased applications).
These local definitions are: \texttt{hd1}, the second
element of the stream \texttt{xs}; \texttt{hd2}, the first element; and
\texttt{tl2}, the exchange of elements of the second tail of \texttt{xs}.
These definitions are then incorporated into the produced stream, where in the
tail we use the abstract constructor \verb|v| of type
\verb|∀ i: Unit. StreamF ·A ·R i ➔ R i|.

In a higher-level language, \texttt{exch} can be written using \emph{nested
  copatterns} (because each step of generation produces a static number of
future observations; course-of-values coiteration permits this number to be
dynamically computed at each step):

{%
  \small%
\begin{verbatim}
head (exch xs)        = head (tail xs)
head (tail (exch xs)) = head xs
tail (tail (exch xs)) = exch (tail (tail xs))
\end{verbatim}%
}%
\section{Coinductive proofs of properties of streams}
\label{sec:proofs}

In this section, we argue that the indexed coiteration scheme enjoyed by
our derived codatatypes is sufficient for giving proofs of many standard
coinductive properties.
We use as an example generalized relations on streams: the lifting of
some relation \verb|Rel: A ➔ A ➔ ★| (over an element type \verb|A: ★|) to a
relation \texttt{StreamRel} over streams in which corresponding elements are
related; instantiation of \verb|Rel| to an equivalence relation produces the
relation of stream bisimilarity (up to equivalence of elements).
The definition of \texttt{StreamRel} is given in Figure
\ref{fig:streamrelf} (note again the use of module parameters, in particular the
import of \texttt{streamf} allowing us to refer by \texttt{Stream} to streams
of elements of type \texttt{A}).
For proofs, we show in Figure \ref{fig:proofs} that if \texttt{Rel} is reflexive,
symmetric, or transitive, then so too is the relation \texttt{StreamRel}.

\begin{figure}
  \centering
  \small
\begin{verbatim}
import utils.

module examples/streamrelf (A: ★) (Rel: A ➔ A ➔ ★).
import streamf ·A.

StreamRelF : (Pair ·Stream ·Stream ➔ ★) ➔ Pair ·Stream ·Stream ➔ ★
= λ R: Pair ·Stream ·Stream ➔ ★. λ p: Pair ·Stream ·Stream.
  [xs = fst p] - [ys = snd p] -
  Pair ·(Rel (head xs) (head ys)) ·(R (intrPair (tail xs) (tail ys))) .

monoStreamRelF : Mono ·(Pair ·Stream ·Stream) ·StreamRelF = <..>

import nu/nu ·(Pair ·Stream ·Stream) ·StreamRelF -monoStreamRelF.

StreamRel : Stream ➔ Stream ➔ ★ = λ xs: Stream. λ ys: Stream. Nu (intrPair xs ys).

headRel : ∀ xs: Stream. ∀ ys: Stream.
          StreamRel xs ys ➔ Rel (head xs) (head ys) = <..>
tailRel : ∀ xs: Stream. ∀ ys: Stream.
          StreamRel xs ys ➔ StreamRel (tail xs) (tail ys) = <..>

unfoldStreamRel : ∀ X: Stream ➔ Stream ➔ ★.
  (∀ R: Stream ➔ Stream ➔ ★.
     (∀ xs: Stream. ∀ ys: Stream. X xs ys ➔ R xs ys) ➔
     ∀ xs: Stream. ∀ ys: Stream. X xs ys ➔
     StreamRelF ·(λ p: Pair ·Stream ·Stream. R (fst p) (snd p)) (intrPair xs ys)) ➔
  ∀ xs: Stream. ∀ ys: Stream. X xs ys ➔ StreamRel xs ys = <..>
\end{verbatim}
  \caption{Definition of generalized relations between streams}
  \label{fig:streamrelf}
\end{figure}

\paragraph{\texttt{StreamRelF} and \texttt{StreamRel}.}
The signature for generalized stream relations, \texttt{StreamRelF}, takes a
type family \verb|R: Pair ·Stream ·Stream ➔ ★| (standing in for recursive
occurrences of the greatest fixpoint of \texttt{StreamRelF} itself) and a pair
\verb|p: Pair ·Stream ·Stream| of the two streams to be related, and in the
body is defined as the type of pairs of proofs that the first elements of the two streams
are related by \texttt{Rel} and proofs that the tails of the streams are related by \texttt{R}.
This type scheme is positive in \texttt{R}, as proven by \texttt{monoStreamRelF}
(definition omitted).
We instantiate the generic codata derivation with these definitions and define
\texttt{StreamRel} as the greatest fixpoint of \texttt{StreamRelF}.
Destructors \texttt{headRel} and \texttt{tailRel} are given (their definitions
are similar to \texttt{head} and \texttt{tail} for streams and so are omitted), as
well as the simplified generator \texttt{unfoldStreamRel}: the properties we
shall prove using \texttt{StreamRel} require only indexed coiteration, so
\texttt{unfoldStreamRel} forgoes the facilities for more advanced generation
schemes (namely, indexed course-of-values coiteration) in order to simplify their proofs.

\begin{figure}
  \centering
  \small
\begin{verbatim}
Reflexive : Π A: ★. (A ➔ A ➔ ★) ➔ ★
= λ A: ★. λ Rel: A ➔ A ➔ ★. ∀ x: A. Rel x x.

Symmetric : Π A: ★. (A ➔ A ➔ ★) ➔ ★
= λ A: ★. λ Rel: A ➔ A ➔ ★. ∀ x: A. ∀ y: A. Rel x y ➔ Rel y x .

Transitive : Π A: ★. (A ➔ A ➔ ★) ➔ ★
= λ A: ★. λ Rel: A ➔ A ➔ ★. ∀ x: A. ∀ y: A. ∀ z: A. Rel x y ➔ Rel y z ➔ Rel x z .

strRefl : Reflexive ·A ·Rel ➔ Reflexive ·Stream ·StreamRel
= λ refl. Λ xs.
  unfoldStreamRel ·(λ xs: Stream. λ ys: Stream. {xs ≃ ys})
    (Λ R. λ ch. Λ xs. Λ ys. λ g.
     intrPair (ρ g - (refl -(head ys))) (ch -(tail xs) -(tail ys) (ρ g - β)))
    -xs -xs β.

strSym : Symmetric ·A ·Rel ➔ Symmetric ·Stream ·StreamRel
= λ sym. Λ xs. Λ ys. λ rel.
  unfoldStreamRel ·(λ ys: Stream. λ xs: Stream. StreamRel xs ys)
    (Λ R. λ ch. Λ ys. Λ xs. λ g.
     intrPair
       (sym -(head xs) -(head ys) (headRel -xs -ys g))
       (ch -(tail ys) -(tail xs) (tailRel -xs -ys g)))
    -ys -xs rel .

strTra : Transitive ·A ·Rel ➔ Transitive ·Stream ·StreamRel
= λ tra. Λ xs. Λ ys. Λ zs. λ rel1. λ rel2.
  [X : Stream ➔ Stream ➔ ★
   = λ xs: Stream. λ zs: Stream.
     ∀ Y: ★. (∀ ys: Stream. StreamRel xs ys ➔ StreamRel ys zs ➔ Y) ➔ Y]
- unfoldStreamRel ·X
    (Λ R. λ ch. Λ xs. Λ zs. λ g.
     g (Λ ys. λ rel1. λ rel2.
          [hd : Rel (head xs) (head zs)
           = tra -(head xs) -(head ys) -(head zs)
                 (headRel -xs -ys rel1) (headRel -ys -zs rel2)]
        - [tl : R   (tail xs) (tail zs)
           = ch -(tail xs) -(tail zs)
               (Λ Y. λ f. f -(tail ys) (tailRel -xs -ys rel1) (tailRel -ys -zs rel2))]
        - intrPair hd tl))
    -xs -zs (Λ Y. λ f. f -ys rel1 rel2) .
\end{verbatim}
  \caption{Coinductive proofs of reflexivity, symmetry, and transitivity for
    \texttt{StreamRel}}
  \label{fig:proofs}
\end{figure}

\paragraph{\texttt{strRefl}.}
To prove reflexivity of \texttt{StreamRel} assuming reflexivity of \texttt{Rel},
we use for the generator the type family of proofs that two streams are equal.
In the body of the coalgebra given to
\texttt{unfoldStreamRel} we have the following obligations: for the head we must
prove \texttt{Rel (head xs) (head ys)}, given by rewriting (with ρ, Figure
\ref{fig:cdle-type-constructs}) by the assumed \verb|g: {xs ≃ ys}| and
using the assumption that \texttt{Rel} is reflexive; for the tail we must prove
\texttt{R (tail xs) (tail ys)}, given by providing the coinductive hypothesis
\texttt{ch} with a proof that \texttt{\string{tail xs ≃ tail ys\string}}.

\paragraph{\texttt{strSym}.}
To prove symmetry of \texttt{StreamRel} assuming symmetry of \texttt{Rel}, we
use for the generator the type family, over streams \texttt{ys} and \texttt{xs},
of proofs that \texttt{xs} is related to \texttt{ys}.
In the body of the coalgebra given to \texttt{unfoldStreamRel} we have the
following obligations: for the head we
must prove \texttt{Rel (head ys) (head xs)}, given by invoking the proof
\texttt{sym} that \texttt{Rel} is symmetric on a proof of \texttt{Rel (head
  xs) (head ys)} which we extract from the head of the assumption
\verb|g: StreamRel xs ys|; for the tail we must prove \texttt{R (tail ys) (tail xs)}, given by
providing the coinductive hypothesis \texttt{ch} with a proof of
\texttt{StreamRel (tail xs) (tail ys)}, extracted from the tail of \texttt{g}.

\paragraph{\texttt{strTra}.}
To prove transitivity of \texttt{StreamRel} assuming transitivity of
\texttt{Rel}, we use for the generator the type family (locally named
\texttt{X}) over streams \texttt{xs} and \texttt{zs} of proofs that
\emph{there exists a \texttt{ys:} \texttt{Stream} such that \texttt{StreamRel xs ys} and
  \texttt{StreamRel ys zs}}.
In the body of the coalgebra given to \texttt{unfoldStreamRel}, we first unpack
the existential \texttt{g} to access these assumptions directly. For the head
(locally named \texttt{hd}) we
must prove \texttt{Rel (head xs) (head zs)}, given by invoking the proof
\texttt{tra} that \texttt{Rel} is transitive on proofs that \texttt{Rel (head
  xs) (head ys)} and that \texttt{Rel (head ys) (head zs)}.
For the tail
(locally named \texttt{tl}) we must prove \texttt{R (tail xs) (tail zs)}, given
by invoking the coinductive hypothesis \texttt{ch} on a proof that their exists
some stream \texttt{ys'} such that \texttt{StreamRel (tail xs) ys'} and \texttt{StreamRel ys' (tail
  zs)}; \texttt{tail ys} is such a stream, with the required proofs extracted
from the tails of the proofs that \texttt{StreamRel xs ys} and \texttt{StreamRel
ys zs}.

\section{Related \& Future Work}
\label{sec:related}
Previous work on lambda encodings of datatypes in Cedille have focused on
inductive datatypes.
In \cite{FS18}, Firsov and Stump generically derive the induction principle for Church-
and Mendler-style lambda encodings; this work was extended by Firsov et al. \cite{FBS18} to
equip datatypes encoded in the Mendler-style with \emph{constant-time}
destructors, and by Firsov et al. \cite{FDJS18} to support
\emph{course-of-values induction} for them.
Our negative result (Section \ref{sec:lambek})
appears to be a real difficulty in adapting the techniques described in
\cite{FBS18, FS18} for efficient constructors of Mendler-style coinductive types, as
the type coercion enabling an efficient recursion scheme used by their (dependent
version of a) Mendler-style algebra 
comes as a consequence of induction, which itself implies \emph{Lambek's lemma}.

To address this difficulty, we use monotone recursive types in which the
introduction and elimination forms require certain \emph{monotonicity witnesses}.
Monotone recursive types of this sort were extensively studied by Matthes
\cite{Mat98b, Mat98} as a way of guaranteeing strong normalization for
extensions of System F that support efficient recursion schemes for inductive datatypes.
These appear to be true extensions: in \cite{SU99}, Spławski and Urzyczyn give
evidence suggesting there can be no efficiency-preserving translation of the
recursion scheme for datatypes in System F.
In contrast, monotone recursive types with the desired computational behavior
are derivable within CDLE (shown by Jenkins and Stump in \cite{JS20}), meaning
no true extensions of CDLE are required (and thus no new meta-theory is needed).
We can therefore define a variant Mendler-style coalgebra equipped with a
zero-cost type coercion whose codomain is the type codatatype being recursively
defined, enabling an efficient corecursion scheme and a constant-time codata
constructor (Figure \ref{fig:nu-3}).

The use of recursive types for constant-time destructors for lambda encodings of
inductive types goes back (at least) to Parigot \cite{Par88} (see also
\cite{Lei89, Men87}).
In \cite{Ge14}, Geuvers analyzed the Church, Scott, and Parigot
(alternatively ``Church-Scott'') method of lambda encodings for both inductive and coinductive data
using a category- and type-theoretic account developed in \cite{Ge92}, and he
similarly uses recursive types to guarantee efficient codata constructors for
the Scott and Parigot encoding.
In comparison to the present work, Parigot-encoded codata supports the
corecursion scheme by requiring recursive occurrences of the codatatype in the
type of its constructors to be embedded in a coproduct type, which requires an
additional case-analysis to access.
This translates to run-time overhead linear in the number of observations made on codata
generated from coiteration alone.
Additionally, \cite{Ge14} does not treat indexed and course-of-values coiteration.

The present work leans upon the account of Mendler-style (co)inductive datatypes
in type theory by Mendler \cite{Me91} and in category theory by Uustalu and Vene
\cite{UV99, UV00}.
The examples of functional programming with codata via different generation
schemes (Section \ref{sec:examples}) come directly from Vene's thesis
\cite{Ven00} (in which solutions were given using ordinary final $F$-coalgebras).
In \cite{UV00}, Uustalu and Vene advocated for the Mendler-style approach as
being a more semantic approach to termination checking of total functional programs.
We agree whole-heartedly, and consider as interesting future work the design of
a high-level language with copatterns using the Mendler-style approach as the basis for
productivity checking, similar to the use of sized types by Abel et al.
\cite{AP13}.
Indeed, Mendler-style recursion schemes were reported by Barthe et
al.~\cite{BFGPU00} as the inspiration for type-based termination checking with
sized types.

\section{Conclusion}
\label{sec:conclusion}
In this paper, we have derived coinductive
types generically using a Mendler-style of
lambda encoding in CDLE, an impredicative Curry-style pure type system.
Codata so derived enjoy direct support for schemes of generation
that are both \emph{expressive} and \emph{efficient}: the \emph{corecursion}
scheme is facilitated by a constant-time type coercion between the concrete and
abstract type of the codata being generated, and the \emph{course-of-values} coiteration
scheme is facilitated by a codata constructor with run-time overhead that is constant in
the number of observations made.
\emph{Indexed coiteration} is also supported, which we demonstrate with examples
of proofs of some standard coinductive properties of streams.
We also showed of our encoding of coinductive datatypes that while the equation
given by one direction of \emph{Lambek's lemma} (which states that final
$F$-coalgebras are isomorphisms) holds by definitional equality, showing the
opposite direction (and thus, coinduction in the sense of ``bisimilarity
implying equality'') appears to require extending CDLE with an extensional
equality type.

\nocite{*}
\bibliographystyle{../eptcsstyle/eptcs}
\bibliography{paper}

\appendix
\section{CDLE: additional term constructors}
\label{sec:appendix-phi}
\begin{figure}[h]
  \centering
  \[
    \begin{array}{c}
      \begin{array}{ccc}
        \infer{
          \Gamma\vdash [t_1,t_2] : \abs{\iota}{x}{T_1}{T_2}
          }{
          \Gamma\vdash t_1 : T_1
          \quad \Gamma\vdash t_2 : [t_1/x]T_2
          \quad |t_1| = |t_2|
          }
        &
          \infer{
          \Gamma\vdash t.1 : T_1
          }{
          \Gamma\vdash t : \abs{\iota}{x}{T_1}{T_2}
          }
        &
          \infer{
          \Gamma\vdash t.2 : [t.1/x]T_2
          }{
          \Gamma\vdash t : \abs{\iota}{x}{T_1}{T_2}
          }
        \end{array}
      \\ \\
      \infer{
      \Gamma \vdash \varphi\ t\ -\ t_1\ \{t_2\} : T
      }{
        \Gamma \vdash t : \{t_1 \simeq t_2\}
        \quad \Gamma \vdash t_1 : T
        \quad \mathit{FV}(t_2) \subseteq \mathit{dom}(\Gamma)
      }
      \\ \\
      \begin{array}{rcl}
        |[ t_1 , t_2 ]| & = & |t_1|
        \\ |t.1| & = & |t|
        \\ |t.2| & = & |t|
        \\ |\varphi\ t\ -\ t_1\ \{t_2\}| & = & |t_2|
      \end{array}
    \end{array}
  \]
  \caption{Dependent intersections and $\varphi$}
  \label{fig:appendix-phi}
\end{figure}

The code listings in Appendix \ref{sec:appendix-derived} detailing the
derivation of type coercions and monotone recursive rely on additional term
constructs for dependent intersections and an additional eliminator $\varphi$
for the equality type.

\paragraph{Dependent intersection}
$\abs{\iota}{x}{T_1}{T_2}$ is the type of terms $t$ which can be assigned
both type $T$ and $[t/x]T'$.
In the annotated language, the introduction form is $[t_1,t_2]$, where $t_1$ has
type $T_1$ and $t_2$ has type $T_2[t_1/x]$.
This is conceptually similar to the introduction rule of a dependent pair, except that $|t_1|$ is
additionally required to be definitionally equal to $t_2$.
This allows the erasure of $[t_1,t_2]$ to be simply $|t_1|$.

Dependent intersections are eliminated with projections $t.1$ and $t.2$: if $t$
has type $\abs{\iota}{x}{T_1}{T_2}$, then $t.1$ has type $T$ and erases to $|t|$
and $t.2$ has type $T_2[t.1/x]$ and also erases to $|t|$.
This can be seen choosing to ``view'' $t$ as either having type $T_1$ or
$T_2[t.1/x]$.

\paragraph{Type coercions by equalities}
The expression $\varphi\ t\ -\ t_1\ \{t_2\}$ (erasing to
$|t_2|$) has type $T$ if $t_1$ has type $T$ and $t$ proves that $t_1$ is equal
to $t_2$. This is similar to the direct computation rule of NuPRL (see Section
2.2 of Allen et al. \cite{ABCEKLM06}).

\section{Derived type constructors}
\label{sec:appendix-derived}
\begin{figure}
  \centering \small
\begin{verbatim}
module utils/cast (I: ★).

Cast : (I ➔ ★) ➔ (I ➔ ★) ➔ ★
= λ A: I ➔ ★. λ B: I ➔ ★. ι f: ∀ i: I. A i ➔ B i. {f ≃ λ x. x}.

intrCast
: ∀ A: I ➔ ★. ∀ B: I ➔ ★.
  ∀ f: ∀ i: I. A i ➔ B i. (∀ i: I. Π a: A i. {f -i a ≃ a}) ➾ Cast ·A ·B
= Λ A. Λ B. Λ f. Λ eq. [ Λ i. λ a. φ (eq -i a) - (f -i a) {a} , β].

elimCast : ∀ A: I ➔ ★. ∀ B: I ➔ ★. Cast ·A ·B ➾ ∀ i: I. A i ➔ B i
= Λ A. Λ B. Λ c. φ c.2 - c.1 {λ x. x}.

-- underscore for anonymous definitions
_ : { intrCast ≃ λ x. x } = β .
_ : { elimCast ≃ λ x. x } = β .

castRefl : ∀ A: I ➔ ★. Cast ·A ·A
= Λ A. intrCast -(Λ _. λ x. x) -(Λ _. λ _. β).

castTrans
: ∀ A: I ➔ ★. ∀ B: I ➔ ★. ∀ C: I ➔ ★. Cast ·A ·B ➾ Cast ·B ·C ➾ Cast ·A ·C
= Λ A. Λ B. Λ C. Λ c1. Λ c2.
  intrCast -(Λ i. λ a. elimCast -c2 -i (elimCast -c1 -i a)) -(Λ _. λ _. β).

_ : { castRefl ≃ λ x. x } = β .
_ : { castTrans ≃ λ x. x } = β .

Mono : ((I ➔ ★) ➔ I ➔ ★) ➔ ★
= λ F: (I ➔ ★) ➔ I ➔ ★. ∀ A: I ➔ ★. ∀ B: I ➔ ★. Cast ·A ·B ➾ Cast ·(F ·A) ·(F ·B).

intrMono
: ∀ F: (I ➔ ★) ➔ I ➔ ★.
  (∀ A: I ➔ ★. ∀ B: I ➔ ★. Cast ·A ·B ➾ Cast ·(F ·A) ·(F ·B)) ➾ Mono ·F
= Λ F. Λ m. Λ A. Λ B. Λ c. intrCast -(elimCast -(m -c)) -(Λ i. λ a. β) .

elimMono : ∀ F: (I ➔ ★) ➔ I ➔ ★. ∀ A: I ➔ ★. ∀ B: I ➔ ★.
  Mono ·F ➾ Cast ·A ·B ➾ ∀ i: I. F ·A i ➔ F ·B i
= Λ F. Λ A. Λ B. Λ cm. Λ c. Λ i. λ f. elimCast -(cm -c) -i f.

_ : { intrMono ≃ λ x. x } = β .
_ : { elimMono ≃ λ x. x } = β .
\end{verbatim}
  \caption{Type coercions and monotonicity witnesses (\texttt{utils/cast.ced})}
  \label{fig:appendix-cast}
\end{figure}
\subsection{Cast}
For an indexing type \verb|I: ★| (given as a module parameter) and type families
\verb|A: I ➔ ★| and \verb|B: I ➔ ★|, type \verb|Cast ·A ·B| is defined as the
dependent intersection type of functions \verb|f: ∀ i: I. A i ➔ B i| and proofs that \verb|f|
is equal to the identity function.

\paragraph{\texttt{intrCast}} takes an argument \verb|f: ∀ i: I. A i ➔ B i| and
a proof that \verb|f| behaves extensionally like an identity function
(\verb|∀ i: I. Π a: A i. {f a ≃ a}|).
In the body of its definition, we introduce a dependent intersection.
For the first argument, we assume \verb|a: A i| and use φ to cast \verb|a| to
the type \verb|B i| of the expression \verb|f -i a| with the equality
\verb|eq -i a| proving these two expressions are equal.
By the erasure of φ, the resulting expression erases to \verb|λ a. a|, so the
proof that this is equal to the identity function (needed for the second
component of the dependent intersection) is trivial, given by β.
As β also erases to \verb|λ x. x|, the two components of the intersection are
definitionally equal, as required.

\paragraph{\texttt{elimCast}} takes an erased cast argument of type
\verb|Cast ·A ·B| and produces a function of type \texttt{∀ i: I. A i ➔ B i}.
Its definition simply uses φ to cast \verb|λ x. x| to the type of \verb|c.1| by
the proof \verb|c.2| that \verb|{ c.1 ≃ λ x. x }|.
This means that \verb|elimCast| itself erases to \verb|λ x. x|.

\paragraph{\texttt{castRefl} and \texttt{castTrans}}
are straight-forward: there is always a cast between two definitionally equal
types, and the composition of identity functions produces an identity function.

\subsection{Mono}
For an indexed type scheme \verb|F: (I ➔ ★) ➔ I ➔ ★|, we define \verb|Mono ·F|
to be the property that any type coercion between indexed type families
\verb|A: I ➔ ★| and \verb|B: I ➔ ★| can be lifted to a coercion between the
type families \verb|F ·A| and \verb|F ·B|.

\paragraph{\texttt{intrMono}} is defined the way it is to make the axiomatic
presentation given in Figure \ref{fig:mono} intelligible.
Its type signature is equal to the type:

\begin{verbatim}
∀ F: (I ➔ ★) ➔ I ➔ ★. Mono ·F ➾ Mono ·F
\end{verbatim}

So, \texttt{intrMono} takes an erased proof \texttt{cm} that there is an \texttt{F}-lifting for
any type coercion, and an erased type coercion \texttt{c: Cast ·A ·B}, and
constructs a new cast from family \texttt{F ·A} to family \texttt{F ·B}.
As all arguments are erased and the definition uses \texttt{intrCast},
\texttt{intrMono} is definitionally equal to \texttt{λ x. x}.

\paragraph{\texttt{elimMono}} takes an erased proof \verb|cm: Mono ·F| and an
erased \verb|c: Cast ·A ·B| and produces a function of type \texttt{∀ i: I. F ·A i
  ➔ F ·B i} by eliminating the cast \texttt{cm -c} of type
\verb|Cast ·(F ·A) ·(F ·B)| on the assumed \verb|f: F ·A i|.
As the definition uses \texttt{elimCast}, \texttt{elimMono} is definitionally
equal to \texttt{λ f. f}.

\begin{figure}
  \centering
  \small
\begin{verbatim}
module utils/rec (I: ★) (F: (I ➔ ★) ➔ I ➔ ★).

import top.
import cast ·I.

Rec : I ➔ ★
= λ i: I. ∀ X: I ➔ ★. Cast ·(F ·X) ·X ➾ X i.

recFold : ∀ X: I ➔ ★. Cast ·(F ·X) ·X ➾ Cast ·Rec ·X
= Λ X. Λ c. intrCast -(Λ i. λ x. x -c) -(Λ i. λ x. β).

recIn : Mono ·F ➾ Cast ·(F ·Rec) ·Rec
= Λ im.
  [f : ∀ i: I. F ·Rec i ➔ Rec i
    = Λ i. λ xs. Λ X. Λ c.
      elimCast -c -i (elimMono -im -(recFold -c) -i xs)]
- intrCast -f -(Λ i. λ xs. β) .

recOut : Mono ·F ➾ Cast ·Rec ·(F ·Rec)
= Λ im.
  [f : ∀ i: I. Rec i ➔ F ·Rec i
    = Λ i. λ x. x ·(F ·Rec) -(im -(recIn -im))]
- intrCast -f -(Λ i. λ x. β) .

roll : Mono ·F ➾ ∀ i: I. F ·Rec i ➔ Rec i
= Λ im. elimCast -(recIn -im) .

unroll : Mono ·F ➾ ∀ i: I. Rec i ➔ F ·Rec i
= Λ im. elimCast -(recOut -im).

_ : {roll   ≃ λ x. x} = β.
_ : {unroll ≃ λ x. x} = β.
\end{verbatim}
  \caption{Monotone recursive types (\texttt{utils/rec.ced})}
  \label{fig:appendix-rec}
\end{figure}

\subsection{\texttt{Rec}}
The definition of \texttt{Rec}, the recursive type former, is described in
\cite{JS20} in detail and follows the proof of Tarski's least fixed-point theorem for
monotone functions over a complete lattice \cite{Tar55}.
Interpreting impredicative quantification as set intersection, read the
definition of \verb|Rec| as the intersection over all type families
\verb|X: I ➔ ★| of all types \verb|X i| such that there exists a type coercion
from families \verb|F ·X| to \verb|X|.

\paragraph{\texttt{recFold}} takes an erased cast \verb|c| from \texttt{F ·X} to
\texttt{X} and produces a cast from \verb|Rec| to \verb|X| by simply giving
\verb|c| to the assumed \texttt{x: Rec} as an erased argument.

\paragraph{\texttt{recIn}} assumes that \verb|F| is monotonic and produces a
cast from families \verb|F ·Rec| to \verb|Rec|.
The function \verb|f| implementing this cast assumes \verb|xs: F ·Rec i| and to
produce an expression of type \verb|Rec| assumes \verb|c: Cast ·(F ·X) ·X|, and
first coerces the type of \verb|xs| to \verb|F ·X i| (by using \verb|recFold|
and the monotonicity witness to lift this to a type coercion from \verb|F ·Rec i|
to \verb|F ·X i|), then uses \verb|c| to coerce this to type \verb|X i|.
x
\paragraph{\texttt{recOut}}
assumes that \verb|F| is monotonic and produces a cast from families
\verb|Rec| to \verb|F ·Rec|.
The function \verb|f| implementing this cast assumes \verb|x: Rec i| and gives
this a type coercion from families \verb|F ·(F ·Rec)| to \verb|F ·Rec| by
combining \verb|recIn| with the assumption of monotonicity.

\paragraph{\texttt{roll} and \texttt{unroll}}
are simply defined by eliminating the casts \verb|recIn| and \verb|recOut|,
resp., so it is immediate that these both erased to \verb|λ x. x|.

\end{document}